\documentclass[11pt,a4paper]{article}
\usepackage[T1]{fontenc}
\usepackage[utf8]{inputenc}
\usepackage{lmodern}
\usepackage[margin=1in]{geometry}
\usepackage{amsmath,amssymb,amsthm,mathtools}
\usepackage{enumitem,microtype}
\usepackage[hidelinks]{hyperref}

\newtheorem{theorem}{Theorem}
\newtheorem{corollary}{Corollary}
\theoremstyle{definition}
\newtheorem{definition}{Definition}
\newcommand{\R}{\mathbb{R}}
\newcommand{\Q}{\mathbb{Q}}
\newcommand{\conv}{\operatorname{conv}}
\newcommand{\interior}{\operatorname{int}}
\newcommand{\PVCP}{\textnormal{PVCP-3D}}
\newcommand{\EPVCP}{\textnormal{EPVCP-3D}}
\newcommand{\PSP}{\textnormal{PSP-3D}}
\newcommand{\BWPSP}{\textnormal{BWPSP-3D}}
\DeclareMathOperator*{\argmax}{arg\,max}
\setlist{itemsep=3pt,topsep=5pt}
\allowdisplaybreaks[2]
\hypersetup{pdftitle={NP-Completeness of the Three-Dimensional Polyhedral Separation Problem},pdfauthor={Guannan Guo}}

\title{NP-Completeness of the Three-Dimensional\protect\\ Polyhedral Separation Problem}
\author{Guannan Guo\\\normalsize Beihang University}
\date{}

\begin{document}
\maketitle

\begin{abstract}
Bone tumor resection planning motivates the following geometric problem. Given tumor points and healthy-bone points in three-dimensional space, find cutting planes that separate the tumor from the surrounding bone. To avoid tumor seeding, the cutting planes must not pass through the tumor. Given a positive integer $k$, select at most $k$ closed halfspaces whose intersection contains all tumor points, while minimizing the total weight of healthy-bone points strictly inside this intersection. We formalize this problem and prove that its threshold decision version is NP-complete. Consequently, the corresponding optimization problem is NP-hard.
\end{abstract}

\section{Introduction}

\begin{definition}[Budgeted Weighted Polyhedral Separation Problem]
\label{def:bwpsp}
An instance of the Budgeted Weighted Polyhedral Separation Problem (BWPSP) consists of:
\begin{enumerate}
\item A finite set $B\subseteq\Q^d$ of healthy-bone points.
\item A finite set $T\subseteq\Q^d$ of tumor points, with $B\cap T=\varnothing$, such that $\conv(T)$ is full-dimensional in $\R^d$. In three dimensions, this means that $T$ contains at least four noncoplanar points.
\item A nonnegative integer $k$, the maximum number of cutting planes.
\item A weight function $w:B\to\Q_{>0}$.
\item A rational loss budget $W\in\Q$.
\end{enumerate}
Rational numbers are encoded by binary numerators and denominators, and integer parameters are also encoded in binary.

The question is whether there exist an integer $r$ with $0\leq r\leq k$ and $r$ closed halfspaces. For each $i\in\{1,\ldots,r\}$, choose $q_i\in\R^d\setminus\{0\}$ and $\beta_i\in\R$, and define
\[
H_i=\{x\in\R^d:q_i^Tx\leq\beta_i\}.
\]
The resulting resection region $P=\bigcap_{i=1}^rH_i$ must contain every $t\in T$ and satisfy
\[
\sum_{b\in B\cap\interior(P)}w(b)\leq W.
\]
For $r=0$, we set $P=\R^d$.
\end{definition}

We append the suffix -3D or -2D to specify the dimension; for example, BWPSP-3D denotes the case $d=3$.

\begin{definition}[Polyhedral Separation Problem]
\label{def:psp}
The Polyhedral Separation Problem (PSP) is the special case of BWPSP in which $w(b)=1$ for every $b\in B$ and $W=0$. Thus every healthy-bone point must be preserved: the chosen halfspaces must contain all tumor points, while the interior of their intersection contains no healthy-bone point.
\end{definition}

\begin{samepage}
\begin{definition}[Three-Dimensional Polytope Vertex Cover Problem]
\label{def:pvcp}
An instance of the Three-Dimensional Polytope Vertex Cover Problem (PVCP-3D) consists of:
\begin{enumerate}
\item A finite set $V\subseteq\Q^3$ consisting of all vertices of a bounded, full-dimensional convex polyhedron.
\item Its edge set $E\subseteq V\times V$, represented by pairs of endpoints in $V$; $E$ is exactly the edge set of $\conv(V)$.
\item A positive integer $k$, the maximum number of vertices that may be selected from $V$.
\end{enumerate}
The question is whether there exists a subset $V'\subseteq V$ with $|V'|\leq k$ such that
\[
\forall (u,v)\in E,\qquad \{u,v\}\cap V'\ne\varnothing.
\]
\end{definition}
\end{samepage}

Das and Goodrich proved that PVCP-3D is NP-complete~\cite{DG97}. We construct a polynomial-time reduction from PVCP-3D to PSP-3D. Since PSP-3D is a special case of BWPSP-3D, both problems are NP-hard. We then prove that BWPSP-3D belongs to NP, establishing NP-completeness for both decision problems.

\section{Related Work}

Automated methods for reducing healthy-bone loss in resection planning have been studied by Hill et al.~\cite{H22} and Romanelli et al.~\cite{R25}. G\'erard~\cite{G16} and Ligthart~\cite{L24} reported the complexity of finite-point polyhedral separation in fixed dimension three, with the number of facets as part of the input, as open in their respective discussions. In dimension two, Edelsbrunner and Preparata~\cite{EP88} gave a polynomial-time algorithm for minimum polygonal separation. Related two-dimensional halfspace maximum-coverage problems and algorithms are also discussed by Badanidiyuru et al.~\cite{BKL11}.

\section{Proof}

We prove a polynomial-time reduction from PVCP-3D to PSP-3D. The NP-completeness of PVCP-3D is established in Theorem~3.4 of Das and Goodrich~\cite{DG97}.

\subsection{Preliminaries}

Given $\PVCP(V_0,E_0,k)$, translate the mean
\[
s=\frac1m\sum_{v\in V_0}v,\qquad m=|V_0|,
\]
to the origin, and denote the translated vertex and edge sets by $V,E$. The instances $\PVCP(V,E,k)$ and $\PVCP(V_0,E_0,k)$ have the same answer.

Let $Q=\conv(V)$. Then $0\in\interior(Q)$, and $Q$ is bounded because $V$ is finite. Let $F$ be the set of all facets of $Q$, and write the outward-facing inequality of each facet $f\in F$ as
\[
\alpha_f^Tq\leq\beta_f.
\]
Since the origin lies strictly inside every facet halfspace, $\beta_f>0$. Setting $t_f=\alpha_f/\beta_f$ gives
\begin{equation}
\label{eq:1}
Q=\{q\in\R^3:t_f^Tq\leq1\text{ for every }f\in F\}.
\end{equation}
Each facet is thus represented by a rational vector $t_f\in\Q^3$. Its computation and encoding length are analyzed in Theorem~\ref{thm:encoding}. Polynomial-bit rational certificates for general feasible cutting planes are established in Theorem~\ref{thm:rational-certificate}.

\subsection{Tumor Points and the Space of Legal Cutting Planes}

We construct $\PSP(T,B,k)$ from $\PVCP(V,E,k)$ as follows. Set
\begin{equation}
\label{eq:2}
T=\{t_f:f\in F\}.
\end{equation}
For each edge $e\in E$, let $f(e),g(e)$ be its two incident facets, and define
\[
\begin{aligned}
c_e&=\frac{t_{f(e)}+t_{g(e)}}2,\\
\gamma&=\min_{\substack{e\in E,\ v\in V\\v\text{ is not an endpoint of }e}}
(1-c_e^Tv),\\
\varepsilon&=\min\left\{\frac14,\frac{\gamma}{4m}\right\}.
\end{aligned}
\]
The healthy-bone point set is
\[
B=\left\{b_e=\frac{c_e}{1-\varepsilon}:e\in E\right\},
\]
and the cutting-plane budget is the same $k$ as in $\PVCP(V,E,k)$.

\begin{theorem}[The origin is an interior point]
\label{thm:origin}
$0\in\interior(\conv(T))$.
\end{theorem}
\begin{proof}
Suppose $0\notin\interior(\conv(T))$. By convex separation, there is a nonzero vector $u$ such that
\[
t_f^Tu\leq0\qquad\text{for every }f\in F.
\]
For every $\lambda\geq0$, we have $t_f^T(\lambda u)\leq0\leq1$. By~\eqref{eq:1}, the entire ray $\{\lambda u:\lambda\geq0\}$ lies in $Q$, contradicting boundedness. Thus $0\in\interior(\conv(T))$; in particular, $T$ contains at least four noncoplanar points.
\end{proof}

\begin{theorem}[Parameter space of legal cutting planes]
\label{thm:cuts}
Every legal halfspace containing $T$ can be normalized as
\begin{equation}
\label{eq:3}
H(q)=\{x\in\R^3:q^Tx\leq1\},
\qquad q\in Q\setminus\{0\}.
\end{equation}
Conversely, every $q\in Q\setminus\{0\}$ defines a legal halfspace containing $T$.
\end{theorem}
\begin{proof}
Consider a halfspace containing $T$,
\[
H=\{x:a^Tx\leq\beta\},\qquad a\ne0.
\]
By Theorem~\ref{thm:origin}, $\conv(T)$ contains an open neighborhood of the origin. Since $H$ contains $\conv(T)$, we have $\beta>0$. Let $q=a/\beta$. Then $H=H(q)$, and $t_f^Tq\leq1$ for every $t_f\in T$, so~\eqref{eq:1} gives $q\in Q\setminus\{0\}$. The converse follows from the same inequalities.
\end{proof}

\subsection{Exposing an Edge}

Every edge of a full-dimensional convex polyhedron in three dimensions belongs to exactly two facets. For an edge $e$, denote these facets by $f(e),g(e)$ and set
\begin{equation}
\label{eq:4}
c_e=\frac{t_{f(e)}+t_{g(e)}}2.
\end{equation}
Here $c_e$ is a three-dimensional vector, unlike the scalar halfspace threshold $\beta_i$.

\begin{theorem}[Edge exposure]
\label{thm:edge-exposure}
For every edge $e$,
\begin{equation}
\label{eq:5}
\{q\in Q:c_e^Tq=1\}=e.
\end{equation}
Consequently, for every vertex $v\in V$,
\begin{equation}
\label{eq:6}
c_e^Tv=
\begin{cases}
1,&v\text{ is an endpoint of }e,\\
<1,&v\text{ is not an endpoint of }e.
\end{cases}
\end{equation}
\end{theorem}
\begin{proof}
For every $q\in Q$, the definition of $Q$ gives $t_{f(e)}^Tq\leq1$ and $t_{g(e)}^Tq\leq1$. Thus
\[
c_e^Tq=\frac{t_{f(e)}^Tq+t_{g(e)}^Tq}{2}=1
\]
if and only if both inequalities hold with equality. This means that $q$ belongs to both facets, whose intersection is precisely $e$. The only vertices on an edge are its two endpoints, proving~\eqref{eq:6}.
\end{proof}

\subsection{Validity of the PSP-3D Instance}

Define the uniform gap
\begin{equation}
\label{eq:7}
\gamma=\min_{\substack{e\in E,\ v\in V\\v\text{ is not an endpoint of }e}}
(1-c_e^Tv).
\end{equation}
Since $Q$ is full-dimensional, $m\geq4$, so for every edge $e$ there is a vertex $v\in V$ that is not an endpoint of $e$. The minimum is therefore over a nonempty finite set. Every term is positive by Theorem~\ref{thm:edge-exposure}, so $\gamma>0$. Choose
\begin{equation}
\label{eq:8}
\varepsilon=\min\left\{\frac14,\frac{\gamma}{4m}\right\}.
\end{equation}
Then
\begin{equation}
\label{eq:9}
0<\varepsilon\leq\frac14,
\qquad \varepsilon<\frac{\gamma}{m}\leq\gamma.
\end{equation}
For each edge $e\in E$ in $\PVCP(V,E,k)$, construct one healthy-bone point in PSP-3D:
\begin{equation}
\label{eq:10}
b_e=\frac{c_e}{1-\varepsilon},
\qquad B=\{b_e:e\in E\}.
\end{equation}
By the definition of preservation and $1-\varepsilon>0$, a normalized halfspace $H(q)$ preserves $b_e$ exactly when
\begin{equation}
\label{eq:11}
q^Tb_e\geq1
\quad\Longleftrightarrow\quad
c_e^Tq\geq1-\varepsilon.
\end{equation}
This connects the two problems:
\begin{enumerate}
\item In PSP-3D, $q^Tb_e\geq1$ means that $b_e$ is not strictly inside $H(q)$ and hence is not lost.
\item In $Q$, $c_e^Tq\geq1-\varepsilon$ selects a thin region near the edge $e$. If $c_e^Tq\geq1$, then $q$ lies on the line segment joining the endpoints of $e$; if $q$ is also a vertex of $Q$, then it must be an endpoint of $e$.
\end{enumerate}

\begin{theorem}[Validity of the PSP-3D instance]
\label{thm:valid}
The instance $\PSP(T,B,k)$ defined by~\eqref{eq:2} and~\eqref{eq:10} is valid: $T$ and $B$ are finite rational point sets, $\conv(T)$ is full-dimensional, the tumor points are pairwise distinct, the healthy-bone points are pairwise distinct, and
\[
B\cap\conv(T)=\varnothing.
\]
\end{theorem}
\begin{proof}
The construction uses finitely many rational operations with nonzero divisors. Thus both point sets are rational and finite. Theorem~\ref{thm:origin} gives full-dimensionality of $\conv(T)$. Two different facets cannot have the same normalized supporting plane, so their vectors $t_f$ are distinct.

Suppose that two constructed healthy-bone points coincide, say $b_e=b_{e'}$. The common scale factor $1/(1-\varepsilon)$ gives $c_e=c_{e'}$. By Theorem~\ref{thm:edge-exposure}, the set of points in $Q$ on which this vector has value $1$ is both $e$ and $e'$, so $e=e'$. Distinct edges therefore give distinct healthy-bone points, and $B$ has no duplicate elements.

Finally, to show $B\cap\conv(T)=\varnothing$, choose an endpoint $v$ of any edge $e$. Because $v\in Q$, every $t\in T$ satisfies $v^Tt\leq1$. Consequently, every $t'\in\conv(T)$ also satisfies $v^Tt'\leq1$. However,
\[
v^Tb_e=\frac{c_e^Tv}{1-\varepsilon}=\frac1{1-\varepsilon}>1.
\]
Thus $b_e\notin\conv(T)$.
\end{proof}

\subsection{The Preservation Sets of Vertex Cuts}

A nonzero point $q\in Q$ has two roles. In PSP-3D it specifies a legal halfspace $H(q)$ containing all tumor points. In PVCP-3D the decision selects a subset of $V$, and every candidate vertex belongs to $Q=\conv(V)$.

For each $q\in Q$, define
\begin{equation}
\label{eq:12}
S(q)=\{e\in E:c_e^Tq\geq1-\varepsilon\}.
\end{equation}
For $v\in V$, let $\delta(v)$ denote the set of edges incident to $v$.

\begin{theorem}[Preservation set of a vertex cut]
\label{thm:vertex}
For every $v\in V$, we have $S(v)=\delta(v)$.
\end{theorem}
\begin{proof}
If $v$ is an endpoint of $e$, then $c_e^Tv=1>1-\varepsilon$, so $e\in S(v)$. Otherwise, by~\eqref{eq:7} and~\eqref{eq:9},
\[
c_e^Tv\leq1-\gamma<1-\varepsilon,
\]
so $e\notin S(v)$.
\end{proof}

This allows us to extend PVCP-3D by permitting the selection of arbitrary points of $Q$.

\begin{definition}[Extended PVCP-3D]
\label{def:epvcp}
An instance of Extended PVCP-3D (EPVCP-3D) consists of:
\begin{enumerate}
\item The rational vertex set $V$ of a bounded convex polyhedron $Q=\conv(V)$, with $0\in\interior(Q)$.
\item The edge set $E$ of $Q$.
\item A positive integer $k$, the maximum number of points that may be selected from $Q$.
\end{enumerate}
The function $S$ is determined by $V,E$ through~\eqref{eq:4}, \eqref{eq:7}, \eqref{eq:8}, and~\eqref{eq:12}. The question is whether there is a set $V'\subseteq Q$, with $|V'|\leq k$, such that every $e\in E$ belongs to $S(q)$ for some $q\in V'$.
\end{definition}

Since $V\subseteq Q$ and $S(v)=\delta(v)$, every feasible PVCP-3D solution is also a feasible EPVCP-3D solution. We next show that a solution $V'\subseteq Q$ can be replaced by a vertex set $V''\subseteq V$ that covers at least the same edges and satisfies $|V''|\leq|V'|$. This establishes equivalence between the two problems on the stated input domain.

\subsection{Vertex Domination}

\begin{theorem}[Vertex domination]
\label{thm:dominance}
For every $q\in Q$, there is a vertex $v_*\in V$ such that
\begin{equation}
\label{eq:13}
S(q)\subseteq S(v_*)=\delta(v_*).
\end{equation}
\end{theorem}
\begin{proof}
If $q=0$, then $S(q)=\varnothing$, and any vertex suffices. Otherwise, since $Q=\conv(V)$, write
\[
q=\sum_{v\in V}\lambda_vv,
\qquad \lambda_v\geq0,
\qquad \sum_{v\in V}\lambda_v=1.
\]
The coefficients need not be unique. Fix any such representation and choose $v_*\in\argmax_{v\in V}\lambda_v$. Since there are $m$ coefficients,
\[
\lambda_{v_*}\geq\frac1m.
\]
Take any $e\in S(q)$, so $c_e^Tq\geq1-\varepsilon$. If $v_*$ were not an endpoint of $e$, then $c_e^Tv_*\leq1-\gamma$, while $c_e^Tv\leq1$ for all other vertices. Hence
\[
\begin{aligned}
c_e^Tq
&=\sum_{v\in V}\lambda_vc_e^Tv\\
&\leq\lambda_{v_*}(1-\gamma)+\sum_{v\ne v_*}\lambda_v\\
&=1-\lambda_{v_*}\gamma\\
&\leq1-\frac{\gamma}{m}\\
&<1-\varepsilon,
\end{aligned}
\]
a contradiction. Therefore $v_*$ is an endpoint of $e$.

The vertex $v_*$ depends only on the fixed convex combination, not on $e$. Thus the same vertex is incident to every edge in $S(q)$, giving $S(q)\subseteq\delta(v_*)$. Theorem~\ref{thm:vertex} completes the proof.
\end{proof}

\begin{theorem}[Equivalence of PVCP-3D and EPVCP-3D]
\label{thm:epvcp}
For inputs satisfying $0\in\interior(\conv(V))$, the instances $\PVCP(V,E,k)$ and $\EPVCP(V,E,k)$ have the same answer and are polynomial-time reducible to each other.
\end{theorem}
\begin{proof}
If $C\subseteq V$ is a feasible solution to $\PVCP(V,E,k)$, then $V\subseteq Q$ and Theorem~\ref{thm:vertex} imply that $C$ is also a feasible solution to $\EPVCP(V,E,k)$.

Conversely, let $C\subseteq Q$ be a feasible solution to $\EPVCP(V,E,k)$. For each $q\in C$, Theorem~\ref{thm:dominance} supplies a vertex $v_q\in V$ with $S(q)\subseteq\delta(v_q)$. Set $C'=\{v_q:q\in C\}$. Then $|C'|\leq|C|\leq k$ and
\[
E=\bigcup_{q\in C}S(q)
\subseteq\bigcup_{v\in C'}\delta(v)
\subseteq E.
\]
Thus $C'$ covers every edge and is feasible for $\PVCP(V,E,k)$. On the stated input domain, both problems have the same input $(V,E,k)$ and the same answer. The identity map therefore gives polynomial-time reductions in both directions.
\end{proof}

\subsection{Correctness in Both Directions}

\begin{theorem}[Equivalence of the constructed instances]
\label{thm:equivalence}
The instance $\PVCP(V_0,E_0,k)$ has a vertex cover of size at most $k$ if and only if $\PSP(T,B,k)$ has at most $k$ legal halfspaces whose intersection contains $T$ and whose interior contains no point of $B$.
\end{theorem}
\begin{proof}
Translation preserves vertex covers, and Theorem~\ref{thm:epvcp} establishes equivalence between $\PVCP(V,E,k)$ and $\EPVCP(V,E,k)$. It therefore suffices to prove equivalence between $\EPVCP(V,E,k)$ and $\PSP(T,B,k)$.

Since $S(0)=\varnothing$, the zero vector may be removed from an EPVCP-3D solution. By Theorem~\ref{thm:cuts}, selecting at most $k$ nonzero points $q_1,\ldots,q_r\in Q$ is equivalent to selecting at most $k$ legal halfspaces $H(q_1),\ldots,H(q_r)$ containing $T$.

Set $P=\bigcap_{i=1}^rH(q_i)$. Since the intersection is finite,
\[
\interior(P)=\bigcap_{i=1}^r\interior(H(q_i))
=\{x\in\R^3:q_i^Tx<1\text{ for every }i\}.
\]
By~\eqref{eq:11} and~\eqref{eq:12}, for every $e\in E$,
\[
\begin{aligned}
e\in\bigcup_{i=1}^rS(q_i)
&\quad\Longleftrightarrow\quad
\exists i\in\{1,\ldots,r\}:c_e^Tq_i\geq1-\varepsilon\\
&\quad\Longleftrightarrow\quad
\exists i\in\{1,\ldots,r\}:\frac{c_e^Tq_i}{1-\varepsilon}\geq1\\
&\quad\Longleftrightarrow\quad
\exists i\in\{1,\ldots,r\}:q_i^Tb_e\geq1\\
&\quad\Longleftrightarrow\quad
b_e\notin\interior(P).
\end{aligned}
\]
Hence all edges are covered exactly when the interior of the intersection contains no healthy-bone point. The two instances have the same answer.
\end{proof}

\subsection{Construction Time and Encoding Length}

\begin{theorem}[Polynomial complexity of the construction]
\label{thm:encoding}
Let $L$ be the total binary encoding length of $\PVCP(V_0,E_0,k)$. Both the running time of the construction and the total encoding length of $\PSP(T,B,k)$ are polynomial in $L$.
\end{theorem}
\begin{proof}
We bound the number of arithmetic operations, the bit lengths of the numbers involved, and the total output length.

\textbf{Step 1: Enumerating all facets.}
The facets of $Q$ can be computed directly from its rational vertices:
\begin{enumerate}
\item Enumerate all noncollinear triples of vertices and compute a plane through each triple using a cross product.
\item Test whether all vertices lie on the same side of that plane. If so, it is a supporting plane.
\item Orient the inequality toward the side containing all vertices, normalize its positive right-hand side to $1$, and merge duplicate normalized facet vectors.
\end{enumerate}
Every facet contains three noncollinear vertices, so none is omitted. Conversely, a supporting plane obtained from three noncollinear vertices intersects $Q$ in a two-dimensional face and therefore defines a facet. There are at most $\binom m3$ triples, each side test scans $m$ vertices, and duplicates can be removed using polynomially many exact comparisons. Thus facet enumeration requires polynomially many arithmetic operations.

The edge set $E$ is obtained by translating the supplied edge set $E_0$. For each edge, testing which facet inequalities are tight at both endpoints identifies $f(e),g(e)$. Compute all $c_e$, take the minimum of at most $|E|m$ candidate gaps to obtain $\gamma$, and then compute $\varepsilon$ and all $b_e$. These operations are also polynomial in number.

\textbf{Step 2: Bounding individual bit lengths.}
Every input coordinate has encoding length at most $L$, and $m\leq L$. Computing the mean $s$ involves at most $m$ rational summands and one division. A common denominator can be formed by multiplying the input denominators; its bit length is bounded by the sum of their bit lengths and is therefore polynomial in $L$.

Translated coordinates, three-dimensional cross products, supporting-plane constants, and normalized vectors $t_f$ each use a fixed number of rational operations on these values. All divisors are nonzero, and the supporting-plane constants used for normalization are positive. Their numerators and denominators have polynomial bit length.

Each $c_e$ and each candidate gap $1-c_e^Tv$ also uses a fixed number of rational operations. Taking a minimum to obtain $\gamma$ does not increase bit length. Equation~\eqref{eq:8} determines $\varepsilon$, and
\[
b_e=\frac{c_e}{1-\varepsilon},
\qquad 1-\varepsilon\geq\frac34>0.
\]
All output coordinates therefore have polynomial bit length. An inverse-polynomial numerical lower bound on $\gamma$ is unnecessary: even a very small rational gap can be computed and output exactly when its numerator and denominator have polynomial bit length.

\textbf{Step 3: Total output length and running time.}
The constructed instance has $|F|$ tumor points, $|E|$ healthy-bone points, and the original budget $k$, with unit weights and loss budget $W=0$. Facet enumeration gives $|F|\leq\binom m3$, and $|E|\leq\binom m2$. Hence the number of points is polynomial, and the preceding coordinate bounds imply polynomial total encoding length.

Integer arithmetic, greatest-common-divisor computation, and comparison are polynomial-time in the lengths of their operands. Thus the entire construction runs in polynomial time in the binary Turing-machine model.
\end{proof}

\subsection{The Polynomial-Time Reduction}

\begin{theorem}[Reduction theorem]
\label{thm:reduction}
A polynomial-time algorithm transforms every valid $\PVCP(V_0,E_0,k)$ into $\PSP(T,B,k)$ such that
\[
\begin{gathered}
\PVCP(V_0,E_0,k)\text{ is a YES instance}\\
\Longleftrightarrow\\
\PSP(T,B,k)\text{ is a YES instance}.
\end{gathered}
\]
\end{theorem}
\begin{proof}
Theorem~\ref{thm:valid} establishes validity of $\PSP(T,B,k)$, Theorem~\ref{thm:equivalence} establishes equivalence of the answers, and Theorem~\ref{thm:encoding} establishes polynomial construction time and output length. Therefore
\[
\boxed{\PVCP\leq_p\PSP.}
\]
\end{proof}

\begin{corollary}
\label{cor:hard}
PSP-3D and BWPSP-3D are NP-hard.
\end{corollary}
\begin{proof}
PVCP-3D is NP-complete~\cite{DG97}, so Theorem~\ref{thm:reduction} makes PSP-3D NP-hard. Assigning unit weights and loss budget $W=0$ converts $\PSP(T,B,k)$ into the equivalent instance $\BWPSP(B,T,k,w,0)$. Thus BWPSP-3D is also NP-hard.
\end{proof}

\subsection{Membership in NP}

\begin{theorem}[Bound on the number of cutting planes]
\label{thm:cut-bound}
Every YES instance of BWPSP-3D has a feasible solution using at most $\min\{k,|B|\}$ cutting planes.
\end{theorem}
\begin{proof}
Consider a feasible resection region $P$, and let $B'=B\setminus\interior(P)$ be the preserved healthy-bone points. If $B'=\varnothing$, choosing no cutting planes has the same loss. Otherwise, assign each $b\in B'$ to one cutting plane satisfying $a_i^Tb\geq\beta_i$, and discard all planes with no assigned point. The number of remaining planes satisfies
\[
r\leq\min\{k,|B|\}.
\]
Each remaining plane has a nonempty assigned set $B_i$. All tumor points remain contained, and every point of $B'$ remains preserved, so the loss does not increase.
\end{proof}

\begin{theorem}[Polynomial-length rational certificates]
\label{thm:rational-certificate}
Let $L$ be the total binary encoding length of a BWPSP-3D instance. Every YES instance has a rational feasible solution using at most $\min\{k,|B|\}$ cutting planes, with total coefficient encoding length polynomial in $L$.
\end{theorem}
\begin{proof}
Use the construction in the proof of Theorem~\ref{thm:cut-bound} to retain $r\leq\min\{k,|B|\}$ planes with assigned sets $B_1,\ldots,B_r$. If $r=0$, the result is immediate. For each retained plane, choose a nonzero coordinate $a_{ij}$ from the three components in vector $a_i$, and divide both $a_i$ and $\beta_i$ by $|a_{ij}|$. The chosen coordinate becomes $\sigma\in\{-1,1\}$. Consequently, the following rational linear system in $(a,\beta)\in\R^4$ is feasible:
\begin{equation}
\label{eq:14}
\begin{cases}
a^Tt-\beta\leq0,&t\in T,\\
-a^Tb+\beta\leq0,&b\in B_i,\\
a_j=\sigma\in\{-1,1\}.&
\end{cases}
\end{equation}
Its coefficients come only from the input coordinates and the constants $0,\pm1$, so its total encoding length is polynomial in $L$. Replace the equality by two inequalities and write its feasible set as
\[
K=\{z\in\R^4:Az\leq h\}.
\]
Let $z_*$ be a point of minimum Euclidean norm in $K$. Such a point exists because $K$ is nonempty and closed and a norm sublevel set is bounded.

The vector $z_*$ lies in the span of the normals of its active constraints. Otherwise, a sufficiently small displacement opposite to its component perpendicular to this span would preserve all active equalities and all remaining finitely many inequalities, while strictly decreasing the norm. Choose linearly independent active rows spanning this space, write them as $A_I$, and denote their right-hand side by $h_I$. The equality $a_j=\sigma\neq 0$ ensures that the active rank is at least one, and it is at most four inferred from the matrix size.

Thus, for some vector $y$,
\[
z_*=A_I^Ty,\qquad A_Iz_*=h_I.
\]
Since the rows of $A_I$ are linearly independent, $A_IA_I^T$ is invertible. Hence
\[
A_IA_I^Ty=h_I
\quad\Longrightarrow\quad
y=(A_IA_I^T)^{-1}h_I.
\]
Substituting into $z_*=A_I^Ty$ gives
\begin{equation}
\label{eq:15}
z_*=A_I^T(A_IA_I^T)^{-1}h_I.
\end{equation}
The matrix $A_IA_I^T$ has order at most four. The determinant formula shows that $z_*$ is rational and that the numerator and denominator of each coordinate have bit length polynomial in $L$. Since $a_j=\sigma\ne0$, this solution defines a genuine halfspace.

Replace each original halfspace by the corresponding rational halfspace. All tumor points remain contained, and every healthy-bone point in each assigned set $B_i$ remains preserved. Positive weights ensure that the loss cannot increase. Since $r\leq|B|$, the total encoding length of the cutting-plane coefficients is polynomial in $L$.
\end{proof}

\begin{theorem}[Membership in NP]
\label{thm:np}
BWPSP-3D belongs to NP.
\end{theorem}
\begin{proof}
By Theorem~\ref{thm:rational-certificate}, a certificate consists of $r$ and the rational coefficients $(a_i,\beta_i)$, with total encoding length polynomial in the input length. The verifier checks input validity,
\[
0\leq r\leq\min\{k,|B|\},\qquad a_i\ne0,
\qquad a_i^Tt\leq\beta_i\quad\text{for every }i\text{ and }t\in T.
\]
For each $b\in B$, it includes $w(b)$ in the loss exactly when $a_i^Tb<\beta_i$ for every $i$, and checks that the total is at most $W$. If $r=0$, the universal condition is vacuous, so all healthy-bone points contribute to the loss, in agreement with $P=\R^3$.

These checks require polynomially many exact rational operations on numbers of polynomial bit length. Every accepted certificate specifies a feasible solution, and every YES instance has such a certificate. Hence BWPSP-3D belongs to NP.
\end{proof}

\begin{corollary}[NP-completeness]
\label{cor:complete}
PSP-3D and BWPSP-3D are NP-complete. The corresponding problem of maximizing the total preserved healthy-bone weight is NP-hard.
\end{corollary}
\begin{proof}
Corollary~\ref{cor:hard} establishes NP-hardness of both decision problems. Theorem~\ref{thm:np} establishes membership in NP for BWPSP-3D, and the same verifier applies to its special case PSP-3D. An exact optimum for preserved healthy-bone weight determines whether the loss is at most $W$, so the optimization problem is NP-hard.
\end{proof}

\end{document}